# Designing Virtual Environments for Social Engagement in Older Adults

A Qualitative Multi-site Study

Tong Bill Xu

Human Centered Design, Cornell University, tx66@cornell.edu

Armin Mostafavi

Human Centered Design, Cornell University, am2492@cornell.edu

Benjamin Kim

Division of Geriatrics and Palliative Medicine, Center on Aging and Behavioral Research, Weill Cornell Medicine, bek4007@med.cornell.edu

Angella Lee

Human Centered Design, Cornell University, al2354@cornell.edu

Walter Boot

Department of Psychology, Florida State University, boot@psy.fsu.edu

Sara Czaja

Division of Geriatrics and Palliative Medicine, Center on Aging and Behavioral Research, Weill Cornell Medicine, sjc7004@med.cornell.edu

Saleh Kalantari

Human Centered Design, Cornell University, sk3268@cornell.edu

**Abstract.** Virtual reality (VR) is increasingly used as a platform for social interaction, including as a means for older adults to maintain engagement. However, there has been limited research to examine the features of social VR that are most relevant to older adults' experiences. The current study was conducted to qualitatively analyze the behavior of older adults in a collaborative VR environment and evaluate aspects of design that affected their engagement outcomes. We paired 36 participants over the age of 60, from three diverse geographic locations, and asked them to interact in collaborative VR modules. Video-based observation methods and thematic analyses were used to study the resulting interactions. The results indicated a strong link between perceived spatial presence in the VR and social engagement, while also highlighting the importance of individual personality and compatibility. The study provides new insights into design guidelines that could improve social VR programs for older adults.

# 1 INTRODUCTION

There is a growing body of research investigating applications of immersive virtual reality (VR) for older adults. Some of this work includes using VR to assist older adults with developing their physical capabilities, such as rehabilitation and gait function [1, 17, 33, 45, 48, 56]. An additional body of VR literature is oriented toward helping older adults with mental illness [69] and cognitive training [14, 22]. One of the more novel areas of study, which has grown particularly relevant during the COVID-19 pandemic, is the use of VR as a means for promoting social interaction and interpersonal engagement among older adults.

Social isolation, defined as having few social relationships or infrequent social contact with others, is a significant concern for older adults worldwide [78]. For many people such isolation can have adverse impacts on mental and physical health [13, 39, 50, 78]. A systematic review of studies on social isolation and loneliness found that older adults are one of the populations most at risk of experiencing these conditions, in part due to declines in mental and physical abilities leading to social withdraw [37]. Age is also associated with reductions in social network size [77], which in turn predicts greater risk for cognitive impairment [59] and decreased wellbeing [56]. In contrast, many studies show that active social activities and meaningful engagement with others are an integral part of successful aging, and that the benefits of such engagement are substantial, including better cognitive health, physical and emotional health, and life satisfaction [54, 76].

Researchers have found that a majority of older adults wish to remain actively engaged in society and involved in meaningful collaborative activities [26, 35, 67]. Understanding how to best support this goal and enhance opportunities for social engagement among older adults is an important challenge, and it is one that is increasingly being taken up by the medical and social services communities. A recent review found that interventions to promote social support for older adults can effectively improve outcomes such as perceptions of loneliness and disengagement [17]. Unfortunately, such interventions confront significant obstacles that may push older adults toward isolation, such as life changes associated with retirement, mobility challenges, illnesses, and ageism in the broader society [44]. This is especially true for older adults who are of lower socio-economic status or who are part of minority and marginalized groups [23].

One of the advantages of VR interventions is that they can allow older adults who would otherwise have limited opportunities to meet with other people to do so from the safety and comfort of their own residences. The popularity of "virtual travel" has grown tremendously during the COVID-19 era, as a way for individuals to "leave" their immediate environment and take part in new experiences in a safe fashion [38, 66, 68, 72]. Today's VR systems are able to provide a highly immersive and realistic impression of being "present" in a digitally rendered or filmed environment, which enhances cognitive engagement [60]. When such experiences are designed to include a social component, they can become a valuable means of connecting with others, either through engaging in shared activities or simple conversation. Studies have found that interactions in VR spaces are perceived as being more "realistic" or more similar to in-person interactions when compared against other types of digital communications [36]. Indeed, the emergence of VR as a technology can be traced back to long-standing aspirations toward the creation of collaborative spaces that erase geographic distance [7].

Several prior studies have evaluated the usability, feasibility, and acceptability of VR by older adults, and have found respectable levels of positive response, engagement, and comfort with the technology when it was carefully introduced to participants [1, 18, 27, 30]. However, our knowledge in this area is still limited, since there have not been many specific design studies evaluating features of the technology that are most relevant to an older population. This is particularly true when it comes to social VR, where research into the mediated interactions that take place on such platforms has focused almost exclusively on younger adults. Such research includes a great deal of work on design strategies for enhancing virtual social engagement in general [28, 45, 64], analyses of communication modes and the dynamics of interactive activities in VR [3, 21, 40, 46, 47], engagement strategies for long-distance couples and families [41, 42, 80], and the psychology of VR self-presentation and avatars [8, 15], among others. Unfortunately, to the best of the current researchers' knowledge, none of this literature on social VR has specifically focused on the needs of older adults. The current study was designed to help fill this gap by providing a qualitative observation of older adults from different geographic locations as they met in an immersive VR environment and engaged in various collaborative activities.

## 2 BACKGROUND AND RESEARCH QUESTIONS

Social VR provides a unique form of engagement compared with other types of computer-mediated interaction. One important feature in social VR is the high level of presence, which is defined as a subjective feeling of "being there" within a virtual environment [53, 73]. The degree of spatial presence is related to how deeply users are cognitively immersed in the environment, as well as their ability to "act there," which is to say, the fluid translation of commands into the environment [19, 29, 31, 57]. In social VR, users can convey both verbal and non-verbal cues via their avatars while maintaining a level of anonymity and privacy, a phenomenon that has sometimes been described as "Avatar-mediated communication" [34, 62]. Another advantage of social VR environments compared to other forms of mediated communication is that it allows for a rich presentation of experiences that are closely akin to those people enjoy face-to-face, such as standing next to each other and looking at photos [37], or the exchange of objects among users [2][4].

A key factor in engaging social VR experience is the ability to feel that others present with you in the virtual environment. This phenomenon has been defined as "social presence" [12]. Researchers have found that higher reported social presence in VR is associated with closer emotional responses to those experienced in the real world [10, 32, 48, 61]. Additionally, social presence has been positively correlated to enjoyment and trust [52]. Despite the recognized importance of maximizing social presence in VR experiences, design guidelines for the process are relatively under-explored. Parsons and colleagues provided a psychological model to describe various levels of social presence, ranging from simply noticing the existence of other persons in the environment to actively engaging with those individuals [55]. Oh and colleagues developed this work further by examining factors that affected the sense of social presence, divided into a variety of immersive, contextual, and psychological factors [52]. In another review study, the researchers found that social presence was affected by the sense of self-embodiment, proxemics (location and distance between participants), symmetric vs. asymmetric interactions, the presence of non-verbal avatar cues (gaze, facial expression, gesture), and the types of activities that participants were engaged in [79]. Beyond these few studies, however, design factors related to the production of social presence have remained somewhat vague and tenuous, with little robust empirical grounding.

In the broader context of studies on social VR, Williamson and colleagues [74] found that specific spatial design factors can have a strong effect in enabling positive social experiences, by mediating activities such as group formation and the sense of personal space during interactions. When socialization is the primary goal of a VR environment it also becomes extremely important to incorporate functions for body language—the ability of avatars to fluidly convey the user's gestures, facial expressions, gaze direction, and other fundamental aspects of non-verbal communication [43]. Data from such prior studies has given us a valuable preliminary understanding of important features of VR environments that may affect social outcomes; however, the majority of these design studies have relied on participant samples that skew toward younger adults, and it is imperative to collect more information about how factors such as embodiment, presence, and engagement are experienced by older users of social VR.

For older adults, intuitive interfaces may be particularly important in promoting the adoption of social VR technology and comfort with using it. Outside of the social context, older adults have been found to prefer "realistic engagement" scenarios in VR that reflect ordinary environments, rather than "gamified" scenarios that bestow users with superhuman abilities to meet challenging tasks [21, 40]. In one notable study, Baker and colleagues [4] conducted workshops with older adults and found that the key drivers of VR technology's acceptance in this population were behavioral anthropomorphism ("the embodied avatars' ability to speak, move, and act in a human-like manner") and translational factors ("how VR technology translates the movements of the ageing body into the virtual environment"). Thus, the careful design of these environments and especially avatars is likely to be crucial in meeting the specific engagement needs of older populations, which in some aspects may diverge from broader gamified development trajectories in the industry.

## 2.1 Research questions

The current study was conducted to study the social interaction *behavior* of older adults during a social VR program, and to relate that behavior to program design features. Such research has been relatively limited in the VR field, especially among older adults. Baumeister and colleagues [6] called for more observational study in this field, arguing that it can provide greater insights than the self-reports, hypothetical scenarios, and questionnaire ratings that are commonly used. The current study was not designed to manipulate or evaluate specific VR environmental variables; rather, we drew from prior research to try to optimize spatial presence and embodiment in the VR, and then evaluated the collaborative responses of participants who interacted with those design features.

This paper makes two main contributions to our understanding of social VR use by older adults. First, drawing on our observation and thematic analysis of the interaction of different participant pairs, we identified the features of the VR environment that promoted or discouraged social engagement. Second, we identified personal characteristics among the older adult users that affected social engagement in the VR environment. Since this is a qualitative and observational study, more research will be needed to empirically test these conclusions and place them on a sound scientific footing. However, the value of the study is that it provides insights into an understudied area, and helps to identify important and sometimes overlooked aspects of older adults' VR experiences that merit strong consideration as the technology becomes more widespread.

Overall, the study was guided by eight research questions that shaped the directions of our observation and analysis:

RQ1. What social interaction factors will be most strongly associated with the sense of social engagement during VR experiences in this older adult population? (E.g., frequency of conversation, shared interests, collaborative activities, duration of silence.)

RQ2. What will be the relationship between VR content engagement and VR social engagement? Will more active, collaborative interaction with the environment enhance or reduce sociability? What will be the impact of VR content changes on ongoing conversations?

RQ3. What technological features of VR will best support social engagement for the older adults? Will realism and immersion-enhancing features, such as walls that cannot be "teleported" through, constrain the social aspects of the experience?

RQ4. How will the social engagement levels of participants be related to the social engagement levels of their virtual peers/partners? Will these engagement levels demonstrate a pair-wise effect?

RQ5. Will it be possible to identify demographic or personal characteristics that have a notable impact on social engagement levels in VR?

RQ6. How will non-verbal cues be used by participants for communication in the social VR environment? (E.g., the proximity of avatars to each other, or the use of emotive animations and "body language.") Will there be specific risks of misleading/mismatching cues due to the mediation of technology in social VR environments?

RQ7. Will certain individuals tend to dominate in social VR conversations, or to be more passive in such conversations? If so, what are the characteristics of such individuals, and how do they feel about being dominant/non-dominant? Will it be possible to identify elements of the VR technology that heighten or reduce such dynamics?

RQ8. How will group decision-making processes play out in the social VR? (E.g., power dynamics, engagement in arguments, handling of disagreements.)

## 3 METHODS

### 3.1 Design of the social VR environment and experimental procedure

Each of the social engagement tasks included in our environment were based on the four quadrants of McGrath's circumplex model for collaboration [44]: (1) creating a variety of ideas and plans; (2) selecting a solution; (3) discussing disagreements; and (4) carrying out the amended solution. We also conducted informal pilot tests with older adults to obtain feedback about potential shortcoming of the environment and issues experienced, and used this feedback to enhance some aspects of the VR content prior to the experiment. The finalized VR design included three modules: Introduction, Travel & Conversation, and Collaborative Tasks. It should be noted that prior to the Introduction section, participants were given basic training on how to use the VR equipment and had avatars created for them by the moderators, who were research assistants at the respective study sites. The avatars were designed to model each participant's real-life appearance and dress, and were confirmed with the participant to be appealing prior to the start of the experiment. The moderators were involved throughout the social VR engagement process (as discussed in the following paragraphs), and provided technical assistance as needed for the participants.

Prior to arrival at the study site, participants were asked to complete an extensive online demographic survey. Upon arriving at their session, the moderator first administrated a series of questionnaires, including the Montreal Cognitive Assessment (MoCA) and measures of mood and attitudes toward VR (see section 3.3). The participants were then fitted with the VR equipment and instructed in its use. During the Introduction portion of our experiment, participants entered a room in the virtual world that contained a large map of the Earth, where they met with their assigned partner. They were told that they would be going on a "trip" together, and were asked to introduce themselves (their name, where they are from, etc.). They were then asked to discuss which locations shown on the map they would like to visit for the upcoming Travel & Conversation portion of the experiment. For participants who had difficulty choosing, they were offered an opportunity to visit adjacent rooms, which had images of representative landmarks for each of the available locations.

After agreeing on an itinerary, the participants entered the Travel & Conversation module, which involved immersive virtual videos of their selected locations. Moderators accompanied the participants on this tour and prompted them if needed to engage in conversation with each other, for example by asking them to explain what the most interesting landmark was at the site. Due to the nature of the video technology, the participants were not able to change their position within the environment, but they were able to look around in different directions from their fixed position. For the final Collaborative Tasks portion of the experiment, the participants were brought into a VR gallery space where they were instructed to work together to identify which of the photos in the gallery were not landmarks they saw during their VR tour, and remove those pictures from the space (they were not graded on accuracy in this task). They were then asked to collaboratively rearrange the remaining photos into a collage, by adjusting their size and positioning on the wall, to create an artistic documentation of their "travels." Figure 1 shows some of the virtual environments that the participants encountered during these modules.

Between each of the three modules, participants were allowed to remove the VR equipment and take a short break. Upon completion of the final module, the participants completed a brief exit survey and interview about their VR experience, and were given a $50 gift card as compensation for their participation.

The respective Institutional Review Boards for each of the three institutions involved in this research approved all study procedures and surveys prior to the start of research activities. Each site had two moderators present during the experiment sessions: one monitored the VR experience of the participant and the other monitored the physical safety of the participant. All sites used a consumer edition of the Oculus Quest 2 head-mounted display and its accompanying handheld controllers, a Blue Yeti USB microphone, and a Sony SRS-RA3000 speaker to hear the audio from the other site. Continuous video of each participant's VR experience was recorded using the Open Broadcaster Software (OBS).

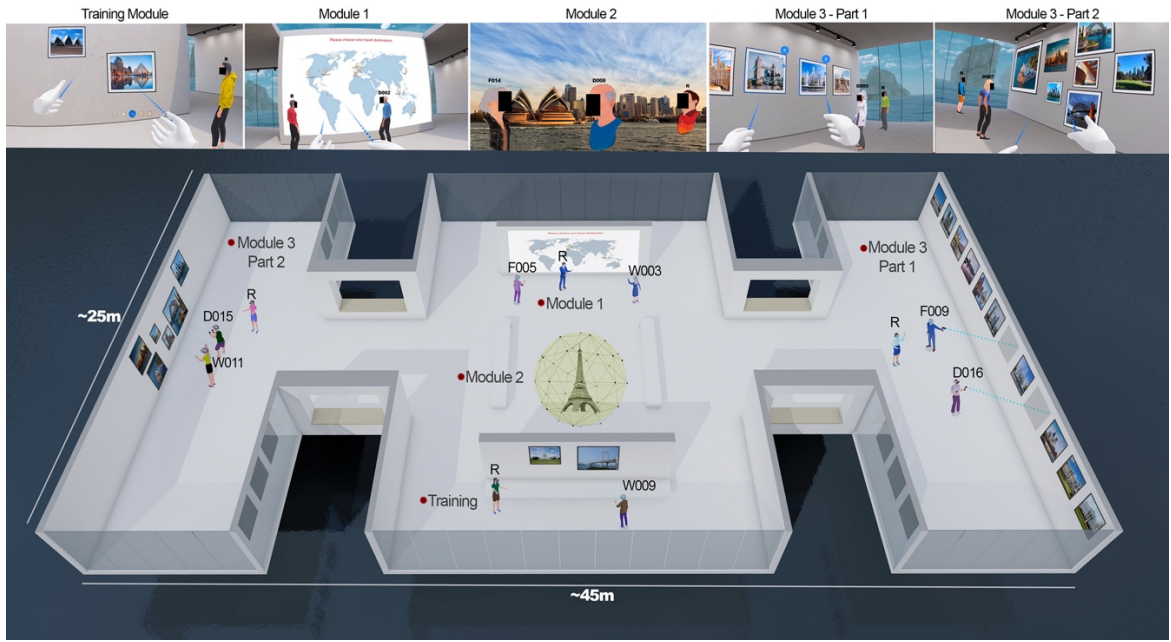

Figure 1: Screenshots from the Participants' View in the VR Modules, Including Training, Introduction (Module 1), Travel & Conversation (Module 2), Collaborative Tasks (Module 3).

## 3.2 Participants

Twelve participants were recruited at each of our three study sites (Ithaca, NY; Tallahassee, FL; and New York City, NY), for a total of 36 participants. We used a convenience sampling method, primarily through the e-mail lists of the participating departments. Each potential participant was engaged personally by the researchers to discuss the study goals, compensation, and potential risks. The inclusion criteria were: (a) at least 60 years of age; (b) able to readily understand written and spoken English; (c) not already a regular user of VR technology. If the participants meet the eligibility criteria and willing to participate after these discussions, participants were provided with an informed-consent document and a demographic questionnaire. We paired each participant with a partner at one of the other sites based on matching schedules (for a total of 18 pairings, six between each site), and collected scheduling availability information so that we could arrange a collaborative VR session for each pair. The participants had an average age of 71 (Standard Deviation = 5.2). The overall sample was mostly Female (72%) and White Caucasian (81%). A wide range of educational backgrounds were represented; more than half of the participants had a professional degree (n=19, 53%), while 5 (14%) had some college education, 10 (28%) had a four-year college degree, and 2 (6%) had a doctorate degree. Table 1 summarizes the participant demographic information and pairings.

Table 1: Characteristics and pairings of participants (n=36)

| Pairs | ID* | Gender | Age | Cognitive Status** | Computer Proficiency | Mobile Proficiency |
|---|---|---|---|---|---|---|
| P1 | D001 | Female | 67 | CI | 28.5 | 33.5 |
|  | W028 | Female | 72 | CI | 30.0 | 39.0 |
| P2 | D002 | Male | 66 | CI | 30.0 | 40.0 |
|  | F022 | Female | 82 | Non-CI | 26.0 | 10.5 |
| P3 | D003 | Female | 61 | Non-CI | 26.0 | 36.5 |
|  | W022 | Female | 70 | CI | 22.0 | 33.5 |
| P4 | D004 | Female | 62 | Non-CI | 30.0 | 18.0 |
|  | F018 | Female | 67 | Non-CI | 27.0 | 39.0 |
| P5 | D006 | Female | 68 | Non-CI | 30.0 | 40.0 |
|  | F004 | Female | 70 | CI | 30.0 | 40.0 |
| P6 | D007 | Female | 66 | Non-CI | 27.0 | 26.0 |
|  | F017 | Male | 75 | Non-CI | 29.5 | 35.5 |
| P7 | D008 | Male | 73 | Non-CI | 26.5 | 34.0 |
|  | F014 | Female | 74 | Non-CI | 27.5 | 35.0 |
| P8 | D013 | Male | 71 | Non-CI | 30.0 | 40.0 |
|  | W008 | Female | 70 | CI | 29.5 | 38.5 |
| P9 | D015 | Male | 67 | Non-CI | 27.0 | 36.0 |
|  | W011 | Female | 69 | Non-CI | 29.5 | 38.0 |
| P10 | D016 | Male | 65 | Non-CI | 24.0 | 32.0 |
|  | F009 | Male | 77 | Non-CI | 30.0 | 40.0 |
| P11 | F010 | Male | 72 | Non-CI | 24.0 | 29.0 |
|  | W012 | Non-binary | 72 | CI | 23.0 | 32.0 |
| P12 | F013 | Female | 72 | Non-CI | 20.5 | 25.5 |
|  | W013 | Female | 74 | Non-CI | 28.5 | 32.5 |
| P13 | W003 | Female | 65 | Non-CI | 24.0 | 31.0 |
|  | F005 | Female | 78 | Non-CI | 27.5 | 36.0 |
| P14 | W006 | Female | 82 | CI | 27.5 | 37.0 |
|  | D005 | Female | 63 | Non-CI | 25.0 | 33.5 |
| P15 | W009 | Female | 67 | Non-CI | 21.5 | 28.5 |
|  | F016 | Female | 75 | Non-CI | 30.0 | 37.0 |
| P16 | W010 | Female | 71 | Non-CI | 21.5 | 21.0 |
|  | F008 | Female | 70 | Non-CI | 30.0 | 36.5 |
| P17 | W021 | Female | 77 | Non-CI | 30.0 | 35.0 |
|  | F012 | Male | 79 | Non-CI | 30.0 | 40.0 |
| P18 | W024 | Female | 66 | CI | 29.0 | 35.0 |
|  | D010 | Female | 73 | Non-CI | 28.0 | 30.5 |

*Participants with ID starting with D were from Ithaca, NY, with W were from New York City, NY, and with F were from Tallahassee, Florida.
**Cognitive status was determined by the MOCA score (<26 Cognitive Impairment, CI; otherwise, Non-CI).

## 3.3 Measures

The current analysis focused primarily on the researchers' observation of the video data from each participant's VR interactions, as well the post-experiment surveys and interviews, which were intended to capture participants' experiences of social interaction. Experiences of physical discomfort with the VR environment were quantified using the Simulator Sickness Questionnaire (SSQ), which included 16 items on a 4-point Likert scale, with higher numbers indicating greater experiences of sickness [33]. The four-item Usability Metric allowed participants to rate the usability of the system using sliders from 0–100, where higher numbers indicate greater usability [20]. The level of participants' Engagement in the VR Experience was measured during the exit survey using a 3-item, 5-point Likert scale developed by the current researchers, with higher scores indicating greater perceived engagement (Cronbach's Alpha = 0.81). To further investigate the usability of the system, the post-survey also included the NASA Task Load Index, with six subscales each rated on a 5-point Likert scale, with higher scores indicating increased perceived task loads [25]. The survey quantified experiences of presence within the virtual environment using the MEC Spatial Questionnaire, which is split into two subscales, Possible Actions and Self-Location [70]. The social aspect of the experiment was assessed using the Social Presence Scale [51] and the Willingness and Likeliness to Reconnect scale [9]. Higher scores in each of these scales indicate, respectively, a greater sense of spatial presence, a greater sense of social presence, and a greater interest in reconnecting with the VR partner. Immediately prior to the VR sessions, participants' cognitive capabilities were evaluated using the *Montreal Cognitive Assessment* [49]. This instrument was used to divide the participants into those with a likely cognitive impairment and those without, using a threshold score of lower than 26 on the instrument as likely-impaired.

## 3.4 Qualitative data analysis

The audio-video recordings of the experiment sessions were transcribed with the aid of the natural language processing algorithm in Adobe Premiere, with manual cleansing and correction. Some markers including participants' laughing moments, participant digital proximities, and changes in the VR content were annotated using the marker tools in the software for further analysis. Participant conversations were coded independently by two researchers, using the following labels: Conversation related to the virtual reality platform and overall VR experience (V); Talking to the moderator (M); Commenting on the task assigned and/or the VR content (E); Experiences from the participants' lives (L); Decision-making processes, either asking or answering (D); Positive emotional expression (P); Comments about non-verbal cues and avatar positioning (N); Technical issues and challenges (T); and Unintelligible segments (C). Examples of each label are shown in Table 2. A third researcher, after watching the recordings and reading the transcripts, went through the labels and refined them together with the two original coders. Finally, a thematic analysis approach [11] was used by all three researchers working together, to identify emerging themes in each label category relevant to the study's research questions.

Table 2: Categorial labels used in the analysis of participant conversations

| Labels | Description | Example |
| --- | --- | --- |
| V | Discussing the virtual reality platform and overall VR experience. | "One of the problems I have is I'm afraid of heights, so I can't look down." |
| M | Talking to the moderator. | "How long are we supposed to do this?" |
| E | Commenting on the task assigned and the VR content. | "I'm going to suggest Madrid." "Beautiful. Looks like cherry blossoms." "I'm making this one bigger." |

| Labels | Description | Example |
|---|---|---|
| L | Experiences from participants' lives. | "I went to a Lutheran school. [name of the school], the high school with the mascot." |
| D | Decision-making process, either asking or answering. | "I don't recall this picture." <br> "Do you want to delete it, or should I delete it?" |
| P | Positive emotion. | "Wow. Blown away by the cherry blossoms. Wow." |
| N | Comments about non-verbal cues and avatar positioning. | "All right, I'm right in front of you now. I know where you are." <br> "Can you hear me?" |
| T | Technical issues and challenges. | "I'm bad with this joystick." <br> "I just can't seem to do it right." |
| C | Unintelligible segments. | [Multiple people talking at the same time] |

## 4 RESULTS

### 4.1 Descriptive statistics

An initial review of quantitative measures found that participants experienced low rates of simulator sickness (M=21.82 out of 235.62; SD=26.69), as well as high usability (M=67.01 out of 100; SD=20.73) and low task workload (M=2.86 out of 7; SD=1.17). Ratings were moderately high for social presence (M=61.21 out of 100; SD=22.00) and spatial presence (possible actions: M=3.70 out of 5; SD=1.17; self-location: M=3.76 out of 5; SD=1.27). Reported levels of engagement were quite high (M=4.18 out of 5.00; SD=0.91) and on average participants had moderately high ratings for likeliness to reconnect with their virtual partner (M=3.69 out of 5; SD=0.79). Table 3 shows the total duration of social interactions, number of words uttered, self-report engagement, self-report likeliness to reconnect, and number of laughs, for each participant pair. Table 4 shows the prevalence of each categorical conversation label.

Table 3: Descriptive statistics for social interactions

| Pair | ID | Number of Words (% total words) | Engagement | Likeliness to Reconnect | Count of Laughs | Duration of Social Interaction (min) |
|---|---|---|---|---|---|---|
| P1 | D001 | 1997 (55.7%) | 2.0 | 2.3 | 25 | 29 |
|    | W028 | 570 (15.9%) | 3.3 | 3.0 |    |    |
| P2 | D002 | 971 (34.2%) | 4.3 | 3.4 | 10 | 27 |
|    | F022 | 1174 (41.3%) | 4.3 | 3.6 |    |    |
| P3 | D003 | 2282 (61.6%) | 2.3 | 2.8 | 11 | 29 |
|    | W022 | 830 (22.4%) | 4.0 | 3.3 |    |    |
| P4 | D004 | 983 (23.9%) | 4.7 | 3.8 | 4 | 29 |
|    | F018 | 2643 (64.2%) | 5.0 | 4.5 |    |    |
| P5 | D006 | 1440 (40.6%) | 4.3 | 3.5 | 27 | 34 |
|    | F004 | 1604 (45.2%) | 4.7 | 4.5 |    |    |
| P6 | D007 | 2397 (45.5%) | 4.3 | 4.1 | 31 | 41 |
|    | F017 | 2072 (39.3%) | 5.0 | 4.5 |    |    |
| P7 | D008 | 1296 (39.4%) | 4.3 | 4.1 | 17 | 36 |
|    | F014 | 1656 (50.3%) | 5.0 | 4.3 |    |    |
| P8 | D013 | 2048 (39.2%) | 4.7 | 2.8 | 13 | 56 |
|    | W008 | 1437 (27.5%) | 5.0 | 5.0 |    |    |
| P9 | D015 | 642 (21.3%) | 3.0 | 3.3 | 8 | 33 |

| Pair | ID | Number of Words (% total words) | Engagement | Likeliness to Reconnect | Count of Laughs | Duration of Social Interaction (min) |
|---|---|---|---|---|---|---|
|  | W011 | 1230 (40.8%) | 5.0 | 4.1 |  |  |
| P10 | D016 | 1337 (35.9%) | 3.7 | 3.5 | 18 | 31 |
|  | F009 | 1309 (35.1%) | 5.0 | 4.3 |  |  |
| P11 | F010 | 1066 (32.7%) | 4.0 | 2.8 | 16 | 25 |
|  | W012 | 1397 (42.9%) | 5.0 | 4.3 |  |  |
| P12 | F013 | 1661 (38.5%) | 2.7 | 3.1 | 19 | 42 |
|  | W013 | 2237 (51.8%) | 4.7 | 4.8 |  |  |
| P13 | W003 | 1000 (38.1%) | 4.0 | 4.0 | 6 | 26 |
|  | F005 | 1002 (38.2%) | 4.0 | 4.1 |  |  |
| P14 | W006 | 1344 (28.6%) | 5.0 | 4.4 | 21 | 36 |
|  | D005 | 2879 (61.3%) | 4.7 | 4.5 |  |  |
| P15 | W009 | 2114 (49.8%) | 4.0 | 3.6 | 9 | 31 |
|  | F016 | 1605 (37.8%) | 5.0 | 4.5 |  |  |
| P16 | W010 | 1822 (44.3%) | 2.3 | 2.3 | 9 | 38 |
|  | F008 | 1604 (39.0%) | 2.3 | 2.0 |  |  |
| P17 | W021 | 2761 (60.7%) | 4.7 | 3.1 | 12 | 38 |
|  | F012 | 921 (20.2%) | 5.0 | 3.9 |  |  |
| P18 | W024 | 797 (15.0%) | 5.0 | 5.0 | 24 | 56 |
|  | D010 | 3499 (65.9%) | 4.0 | 3.0 |  |  |
| Mean |  | 1600 | 4.2 | 3.7 | 16 | 35 |

Table 4: Number of words in each categorical label and percentage of interaction based on number of words in each categorical label divided by total number of words.

| Pairs | M | V | E | L | D | P | N | T | C |
|---|---|---|---|---|---|---|---|---|---|
| P1 | 1628 (45.4%) | 64 (1.8%) | 1115 (31.1%) | 68 (1.9%) | 560 (15.6%) | 144 (4.0%) | 75 (2.1%) | 399 (11.1%) | 88 (2.5%) |
| P2 | 317 (11.2%) | 18 (0.6%) | 1420 (50.0%) | 161 (5.7%) | 1254 (44.2%) | 207 (7.3%) | 0 (0.0%) | 186 (6.5%) | 110 (3.9%) |
| P3 | 45 (1.2%) | 91 (2.5%) | 2793 (75.4%) | 189 (5.1%) | 1004 (27.1%) | 397 (10.7%) | 622 (16.8%) | 942 (25.4%) | 41 (1.1%) |
| P4 | 573 (13.9%) | 0 (0.0%) | 3074 (74.7%) | 354 (8.6%) | 1586 (38.5%) | 465 (11.3%) | 103 (2.5%) | 426 (10.3%) | 93 (2.3%) |
| P5 | 29 (0.8%) | 28 (0.8%) | 2228 (62.8%) | 637 (18.0%) | 853 (24.1%) | 24 (0.7%) | 55 (1.6%) | 122 (3.4%) | 26 (0.7%) |
| P6 | 1 (0.0%) | 86 (1.6%) | 2754 (52.3%) | 910 (17.3%) | 2020 (38.4%) | 22 (0.4%) | 213 (4.0%) | 219 (4.2%) | 39 (0.7%) |
| P7 | 120 (3.6%) | 28 (0.9%) | 1805 (54.8%) | 600 (18.2%) | 1100 (33.4%) | 59 (1.8%) | 78 (2.4%) | 117 (3.6%) | 62 (1.9%) |
| P8 | 174 (3.3%) | 157 (3.0%) | 3307 (63.3%) | 289 (5.5%) | 936 (17.9%) | 11 (0.2%) | 26 (0.5%) | 528 (10.1%) | 15 (0.3%) |
| P9 | 60 (2.0%) | 0 (0.0%) | 1456 (48.3%) | 155 (5.1%) | 351 (11.6%) | 18 (0.6%) | 16 (0.5%) | 177 (5.9%) | 309 (10.3%) |
| P10 | 0 (0.0%) | 89 (2.4%) | 1938 (52.0%) | 408 (10.9%) | 969 (26.0%) | 16 (0.4%) | 0 (0.0%) | 76 (2.0%) | 0 (0.0%) |
| P11 | 0 (0.0%) | 101 (3.1%) | 2162 (66.4%) | 136 (4.2%) | 489 (15.0%) | 65 (2.0%) | 74 (2.3%) | 319 (9.8%) | 0 (0.0%) |

| Pairs | M | V | E | L | D | P | N | T | C |
|---|---|---|---|---|---|---|---|---|---|
| P12 | 119 | 142 | 2250 | 862 | 1777 | 96 | 84 | 243 | 108 |
|  | (2.8%) | (3.3%) | (52.1%) | (20.0%) | (41.2%) | (2.2%) | (1.9%) | (5.6%) | (2.5%) |
| P13 | 157 | 0 | 1572 | 80 | 832 | 64 | 33 | 43 | 107 |
|  | (6.0%) | (0.0%) | (59.9%) | (3.0%) | (31.7%) | (2.4%) | (1.3%) | (1.6%) | (4.1%) |
| P14 | 162 | 62 | 3078 | 584 | 1550 | 943 | 126 | 172 | 36 |
|  | (3.4%) | (1.3%) | (65.5%) | (12.4%) | (33.0%) | (20.1%) | (2.7%) | (3.7%) | (0.8%) |
| P15 | 306 | 89 | 3173 | 264 | 1804 | 171 | 73 | 236 | 27 |
|  | (7.2%) | (2.1%) | (74.8%) | (6.2%) | (42.5%) | (4.0%) | (1.7%) | (5.6%) | (0.6%) |
| P16 | 108 | 9 | 3002 | 44 | 1968 | 312 | 9 | 124 | 51 |
|  | (2.6%) | (0.2%) | (73.0%) | (1.1%) | (47.9%) | (7.6%) | (0.2%) | (3.0%) | (1.2%) |
| P17 | 377 | 0 | 3049 | 51 | 2123 | 150 | 0 | 468 | 143 |
|  | (8.3%) | (0.0%) | (67.0%) | (1.1%) | (46.7%) | (3.3%) | (0.0%) | (10.3%) | (3.1%) |
| P18 | 358 | 51 | 3394 | 361 | 2243 | 552 | 26 | 228 | 69 |
|  | (6.7%) | (1.0%) | (63.9%) | (6.8%) | (42.2%) | (10.4%) | (0.5%) | (4.3%) | (1.3%) |
| Mean | 251.9 | 56.4 | 2420.6 | 341.8 | 1301.1 | 206.4 | 89.6 | 279.2 | 73.6 |
|  | (6.3%) | (1.4%) | (61.0%) | (8.6%) | (32.8%) | (5.2%) | (2.3%) | (7.0%) | (1.9%) |

### 4.2 Social interaction factors associated with engagement (*RQ1*)

In the pairing P1 (Participants D001 and W028), Participant D001 frequently attempted to initiate conversations with the Participant W028 with little response, leading to very limited conversation between the pair. This can be attributed to a lack of shared interests between the two participants and disengagement from Participant W028, which can be noticed in their low count of personal life content ("L") and comments between the pair related to the tasks and VR content ("E"). Alternatively, the pairing P6 (Participants D007 and F017) experienced a much higher degree of social engagement during their VR experience. Social integration factors most strongly related to this sense of social engagement for P6 were the high frequency of personal conversation, indicated by the high count of personal life content ("L"). This pairing also demonstrated patience when the other participant experienced a technical issue, as seen when D007 said "Don't worry about it, [name of their peer]. This is all new." This pair also showed a high count of laughter during their experiment, and a high degree of involvement in the decision-making process during the experiment ("D"). It should also be noted that P6 had such a strong sense of social engagement during their VR experiment that Participant F017 indicated an interest in exchanging information with Participant D007 after the experiment. The most important factor found here was shared interest, which contributed to overall frequency of conversation, engagement in the tasks, frequent instances of positive emotion during conversation, and patience when the other participant was experiencing an issue during the experiment.

### 4.3 The relationship between VR content engagement and VR social engagement (*RQ2*)

Much of the conversation during the experiment, nearly two-thirds of all lines, was about the virtual reality system and its contents. The topics of these conversations were heavily influenced by the content/modules. In *Module 1*, conversation happened at the beginning when participants entered the "Map Room." Participants were interested in VR overall and positive about the room we designed, commenting it to be "extremely modern, which is fantastic." The conversation then moved to the task, deciding the destination of their VR travel. We observed two main patterns in this phase. In some sessions, both participants gave several options and found their common interest, as mentioned by one of them "I'm just choosing three so that we can agree on one." Another strategy, when their peers just gave one option, was to simply agree with them, sometimes with their own reasons to go there ("I'm so intrigued by Australia too, where my daughter had her overseas classroom training"). Our participants rarely disagreed with their peers, even if they were not interested in destinations as we found later in our follow-up interviews, except when their peers proposed an option that was not on our list of available sites.

In *Module 2*, conversations were often triggered by the changes of scenes in the videos (Figure 2). The comments ranged from simple compliments ("Oh, Boy. Beautiful, beautiful pool." "Oh, my goodness. This is beautiful.") to descriptive ("It's bright and sunny here."), and occasionally, they would talk about what they want to do in the place shown in the VR ("It looks like a good place to have lunch." "I'm loving this. I really feel like I'm in Australia. Makes me really want to go sunbathing."). Some of the older adult participants had previously been to, or lived in, the chosen destinations in the past, often decades ago before retirement, and they would compare what they have experienced in the past with what has been filmed, sharing their memory with their peers: "I remember when it was being built . . . . I don't know how long ago that was, [nearly] 30 years." Sometimes they also tried to respond to other people in the videos, waving their hands or greeting them, with additional comments like, "I wonder if these people knew they were being photographed." Since there was no real task for Module 2 other than enjoying the videos, all conversations were about the content presented or the participants' own lives. It is also worth noting that multiple participants mentioned the perception of height in some of the VR videos, which caused a certain level of anxiety ("I can't look down. I'm afraid of heights.") or safety concerns ("I feel like I could fall down there.").

In *Module 3*, most conversations were about the tasks and decision-making. The majority of the lines here fell into three categories: Fact, Opinion, and Action. The first category, Fact, occurred during the first half of the module when participants were asked to remove the irrelevant pictures that were not part of their trips. While there were correct answers to the questions (the pictures presented were actual photos from the destinations), the participants were usually not very confident about their memories, partially due to the fact that many had spent more time talking with each other than really watching the videos during Module 2. Usually before making a decision by themselves, they would provide their reasons and ask for opinions or confirmation from their peers ("It doesn't look like Asian people. What do [you] think?"), while their peers would respond either with additional evidence ("This is a church. It doesn't look like it would be in Tokyo.") or simply based on their own memory ("I don't remember that either"). Again, we observed very few arguments, and most participants seemed more concerned with maintaining a good relationship than having their way in the task completion. The second category, Opinion, happened mostly during the second half of Module 3 where we asked participants to make a collage from the pictures. Here the participants were more confident to give others their preferences ("Four from the left. This is my favorite."), but they still showed little proclivity to argument. The last category of statements, Action, were about informing the partner what they were doing, or giving consent to an activity ("I'm trying to expand this picture because it's so beautiful." "You can take the frame away from the other.").

Conversations that had little relation to the VR content took up about one-tenth of the lines. In Module 1, participants frequently talked about where they grew up, where they had previously lived, and where they had previously traveled. These sections of conversation tended to be extended if the participant pairs happened to find some shared experiences, such as both growing up in New York City. Similarly, in Module 2 it was common for conversations to focus on past life experiences ("I grew up in New York City, so we've got a lot of urban experience." "I was in Sacramento during my high school years. When skateboarding first started out."). In some cases, participants discussed even more divergent topics, such as book clubs and literary interests. Occasionally, these "irrelevant" discussions continued into Module 3, but they were less common there as most of the dialogue shifted to focus on the task completion.

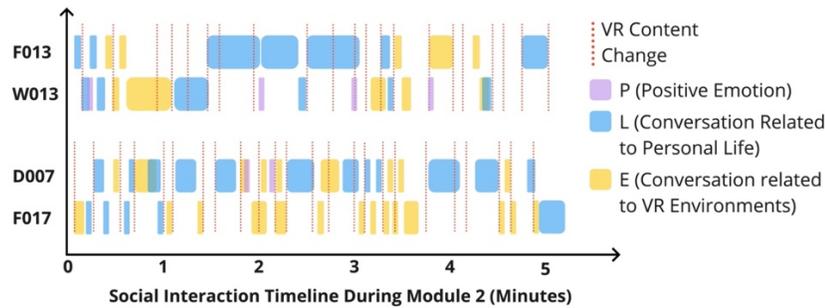

Figure 2: Impact of content change on different types of conversation during the social interaction in Module 2 among participants F013 and W013 (Pair 12) and D007 and F017 (Pair 6)

### 4.4 Technological features of VR that support social engagement (*RQ3*)

As participants navigated around the virtual environment, they encountered various features, bugs, and the limitations of virtual reality. One unexpected outcome was that some participants found they were able to access certain control functions that we had not intended to enable. Our design made use of commercial software, and not all of the available features were incorporated into the experiment. Some participants discovered the ability to open a hidden menu, and one discovered a pen option, which they could use to create a 3D drawing. The participant promptly used the pen to add their partner's name to their final collage. When participants discovered new or unexpected features, they reacted in varied ways, ranging from surprise, to curiosity, to frustration. One who discovered the menu faced it with frustration, because they were unable to figure out how to close it from their screen. In Module 3, some participants inadvertently teleported into the water surrounding the gallery space, and one declared, some incongruously, "It's cold!"

It should also be noted that while the sound decays in a real environment, this was not the case for our experiment as we used a background Zoom call for participants' conversations. Hence, the perception of interpersonal distances was affected by the fact that the social interaction was not dependent on the participants' audio range. If we consider the proximities introduced by Hall [24], then the observed interpersonal distances between the pairs for our study were less frequently in the "personal" range (0.5m–1.2m) and were more frequently in the "social" range (1.2m–3.6m) and "public" range (>3.6m). The relatively low level of computer proficiency and controller familiarity among our participants may have also affected these interpersonal distances, causing participants to be more cautious with their positioning. Overall, our environment had a relatively high level of "unrealistic" features, such as the ability to teleport and the lack of audio drop-off over distance, which may have led to a somewhat greater sense of separation between the social component and the VR component. This did not, however, appear to negatively impact the level of social engagement, and we speculate that it may have contributed to high incidence of "off-topic" conversational development that was not associated with specific VR content and tasks (to draw any robust conclusions about this relationship, a comparative design study would be needed).

Our participants also encountered and talked about many challenges during the experiment, contributing to about 6% of the conversational lines. Most of these challenges were caused or partially caused by the design of the VR, including both software and the hardware components. Most older adults suffer from a certain level of cognitive and/or physical decline. For example, participant D013 asked, "Which button are you clicking when you're on the X?" It was hard for many of the older adult participants to learn and memorize the VR controls within a short timeframe, and visual prompts were not available for reference. (This is highly unusual considering even modern action games would have an option to show control schemes on screen.) This deficiency in our environment was remarked upon by multiple participants. Furthermore, many older adults suffer from decline in fine motor skills, which can make it harder for them to perform the required actions, such as pointing the controller at a very small button on screen and pressing the selection key, precisely and correctly. The system was not able to handle concurrent button pressing events, and many of the participants unintentionally pressed the "grip" buttons on the handle of the controllers by holding them too tightly, which stopped the system from responding to the trigger buttons for interactions. This sometimes resulted in anxiety or frustration, leading to holding the controllers even more tightly, creating a negative feedback loop.

Joystick control was another common technical concern that emerged. In the controller system that we used, pushing the joystick precisely forward would initiate a teleportation action, while a horizontal push would result in a rotation within the VR environment. Participants frequently mixed up these commands, or initiated both at the same time, leading to chaotic movement. Mapping two different control schemes to very similar joystick push actions should not be recommended. The pinnacle of the confusion occurred when participants tried to change the size of the pictures in the VR, which required: (1) pointing controllers on both hands at the picture steadily; (2) pressing trigger buttons on both controllers at the same time; (3) moving the controllers closer to, or away from, each other while holding the trigger buttons; and (4) avoiding triggering the "hold" buttons on the handle during the whole process.

Additionally, there was a frequent mismatch between the physical and the VR environments in terms of the height of the users. Most of our older users preferred to sit during the experiment to minimize physical workload, however, as one participant mentioned, "I look down, I feel like I'm about six feet off the ground." Another major limitation was the VR headset's lack of compatibility with some types of glasses, which are extremely common necessities for older adults. Many of our participants discussed this, for example:

> W013: "Are you wearing glasses under your headset?"
> F013: "Yes, I am."
> W013: "Is that working out for you? Do you have bifocals or regular glasses?"
> F013: "I have a progressive . . . . And that's working under the headset."
> W013: "Okay . . . unfortunately, I have glasses usually, but I can't wear them with the headset . . . . But I'll count on you to see better than I in that case."

Sometimes experimenters had to intervene when such technical issues became a major obstacle for participant's engagement with the VR. This may in some cases have hampered the fluid development of social interactions between our participants.

### 4.5 Peer effects on social engagement (*RQ4*)

We found that the social engagement of participants in the VR social environments affected the social engagement of their virtual reality partner. For example, in the pairing P1 (Participants D001 and W028), we observed that Participant W028 spoke more with the researcher than their paired participant. In addition, Participant W028 said in their exit interview that "The imagery was great, but I can't say it was successfully social interaction." In response to the question, "Which part could be most improved and why?", Participant D001 replied "Oh my partner. I don't want to come down hard on this guy because this is a personality thing not the project, hope he's not sensitive. It's kind of like when you're teaching a class, like a kindergarten class, and you have to explain how you have to raise your hand to use the restroom and some kids just bolt out the door and some people are around, so [the] lack of interaction was very frustrating for me." Interestingly, despite Participant W028's limited social presence during the experiment, she demonstrated high spatial presence, as indicated by her response in their exit interview that "the imagery was great" and frequent comments related to how immersive her experience was. In contrast, there were instances where the social engagement of participants in the VR social environments affected the social engagement of the virtual peers in a more positive manner. This was observed in P4 (Participants D004 and F018) where Participant D004 revealed that they had "zero experience with these controllers" and Participant F018 responded by saying, "'I'm an advocate, and I've only had one other experience with controllers. You will be fine. You will be just fine." This encouragement at the onset of the experiment likely led to increased social engagement between the pair, as indicated by the higher counts of personal life content ("L") and instances of laugher during this experiment. As we can see from the data, social engagement is usually mutual.

**4.6 The effect of personal characteristics on social engagement in VR (*RQ5*)**

Among the various demographic variables that we examined, we found that the profession of the participant and shared interests and experiences had the most notable impact on social engagement. For example, Participant D013 worked in IT support. This participant led to instances of him guiding his paired participant in a particularly efficient and patient manner, which can be noted when he said, "Okay, so you're going to go put your pointer up and I'll picture again and you'll see the icons and the one on the bottom is looks like it. There you go." In P6, (Participants D007 and F017), Participant D007 revealed they were a nurse and also had the highest count of laughter during the experiment, the highest count of personal life content ("L"), and several instances of composure and explanation as seen when Participant D007 said "Yeah, don't worry about it . . . you can use your left hand, which I didn't do." Participant W009 revealed they were a teacher (typically described as "assertive", "caring", and "focused"), and had the high count of decision-making conversations; a high count of comments related to the tasks and VR content, and a tendency to dictate the direction of the conversation as seen in the following examples:

> [During Module 1]
> W009: "I'd like to be in Sydney, Australia. I've never been to that part of the world and I would love to go exploring there."
> F016: "My husband was when he was in Vietnam, when finals are in Sydney and he said it was a great city."
> W009: "Yeah, I wouldn't want to be in Tokyo. It's very crowded."
> [During Module 3]
> W009: "You don't. Yeah. And this one is also from . . . . I can move this one over here. It's not a like a collage. I would move it down a little bit."
> F016: "Oh, okay. You want to make all right. You want to make an actual [collage]. Well, I can do that. Okay."

Another characteristic we observed to have a notable impact on social engagement in VR was shared experiences between participants. For instance, in P12 (Participants F013 and W013), both participants found that they were involved in book clubs and enjoyed reading for leisure. P12 had the highest percentage of personal life content ("L"), which may have also contributed to their successful collaboration and shared engagement, as indicated by their high counts of decision-making content ("D"). In P6 (Participants D007 and F017), the participants discovered that they had both visited San

Francisco before, which may have also contributed to their successful social engagement in VR as indicated by their high counts of conversations related to their personal life ("L") and frequent instances of laughter during their experiment.

### 4.7 Non-verbal cues used by participants for communication in social VR environments (*RQ6*)

Throughout the experiments, there were several references to participants' virtual avatars and bodies, which suggests their importance to participants. These references can be categorized as awareness of one's own avatar vs. awareness of others' avatars. Beginning with awareness of self, participants mentioned being wary of hitting others, despite there being no consequences to colliding/overlapping avatars in the virtual environment. One participant stated, "I'm going to make myself go to this room so that I—Oh, I think I just hit you." Participants sometimes asked whether the other participant could see them, and also commented when they could see the other participant. They made comments such as, "Okay, I see you. You're to my left.", and "All right, I'm right in front of you now." They also made judgements about the realism of the avatars they were seeing. For example, "Oh my, she's floating!" and "Uh oh, is she crunched up?" These statements were in response to avatars that did not seem realistic, because they were not on the ground or had limbs in unnatural configurations, respectively. When participants did look realistic, they appreciated it. One participant said, "You're wearing a yellow sweater. I must be looking at you." These findings show the importance of allow participants to customize their avatars and enhancing avatar realism.

Several allusions were also made to participants' distance from each other and location within the environment. In some cases, this occurred when pairs were seeking one another's location because they could hear each other but could not visually locate one another; for instance:

> W021: Oh, I'm just behind you. I'll try to get next to you. I'm not far behind you.
> F012: Come over here.
> W021: Where are you? I'm over here."

They have also been surprised to find each other at a personal distance "Look, we are right next to each other. We're right next to each other." "This is funny!" Moreover, the participant commented about the others' location accuracy, "If you can come over here closer to me, then you will not be between those two pictures. Like you are kind of stuck between them right now." All of these references to others' virtual avatars/bodies, location, and distances suggest that our participants were trying to transfer their knowledge about physical social interaction in the real world into the virtual reality. They were immersed in the realistic aspects of VR, and felt confused or amused by unnatural presentations.

### 4.8 Dominance, power dynamics, and collaborative decision-making in social VR (*RQ7* & *RQ8*)

Among the pairs in this study, three types of interactions tended to lead to dominance in the conversation. The first was task-oriented and was especially prevalent in Module 3. For example, one participant would lead the decision-making process of which pictures were removed from the wall, as identified by the frequent use of imperative statements or sometimes, questions about the other participant's preferences. These task-dominant participants could be identified through phrases such as, "We have to move these two photos" and "You got it. All right, so move the other one. Move this ridge one right now. Just move it to this wall, actually. All right. Now we're supposed to make like, a collage." Examples of questions included "I think it's kind of cooler without the frames. What do you think?" The second type of dominance interaction was exemplified by pairs with significant skill gaps in understanding the virtual reality controls, including how to teleport and manipulate images. Among these pairs, the skilled participant would either teach the unskilled participant or complete the tasks themselves with (at most) narration and verbal input from the unskilled participant. In one pair, the unskilled participant said to the skilled participant, "You do the heavy lifting," demonstrating a relinquishing of control. The third type of dominance interaction was related to the initiation of conversations. Sometimes, pairs would feature one participant who tolerated silence less than the other participant. These participants would essentially always break silences by asking questions, such as "When was this made? Was this before the fire?" and "[Participant name], how are you doing?" These interactions demonstrate that dominance can take different forms, based on the specific roles that people are inclined to take in social interactions. Participants may be dominant in some ways, but not in others. Some examples of social interaction timelines with pairs with dominant participants and paired with a balanced dominancy in Modules 2 and 3 were illustrated in Figure 3.

Examining the decision-making processes in the virtual environment, this study found mostly that participants moved quickly toward agreement. The cycle of events involved asking, sharing, and agreeing or disagreeing, until unanimity was reached. Some pairs lacked either asking or sharing, but not both. In most pairs, decisions were made fairly quickly, with only a couple instances of disagreement. Most typically, one participant would ask the other participant for their opinion: "If we were to make one picture huge, bigger than it is, which one would you choose?" Next, participants would share something about their lives or from their memory to add context, and present a choice: "No, I'm not a world traveler, too, but I kind of always wanted to go to Singapore." Finally, the participant who brought up the decision could choose to agree or disagree: "No, this is not Tokyo. We didn't go over that bridge." In this study, decision-making was concentrated in Modules 1 and 3 and did not entail many lines of speech, perhaps because participants had little personal investment in the decisions being made.

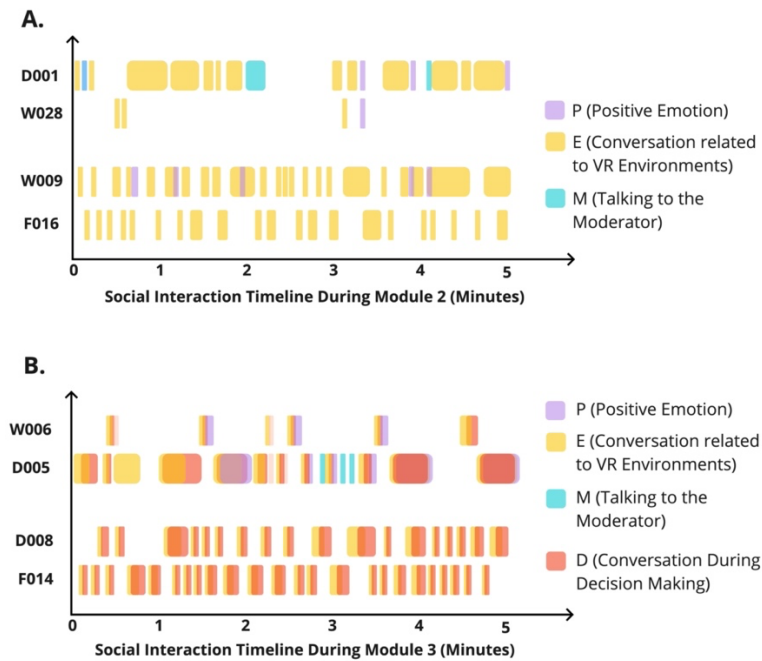

Figure 3: A) Example of conversation dominancy during social interaction in Module 2 observed in the Pair 1 (D001 and W028) and a balanced dominancy in the Pair 15 (W009 and F016). B) Example of different power dynamics during the decision-making

collaborative tasks in Module 3 observed in Pair 14 (W006 and D005; dominant by D005) as compared to Pair 7 (D008 and F014; balanced conversation during decision making).

## 5 DISCUSSION AND CONCLUSION

This paper reports a behavioral social interaction analysis of 18 pairs of older adults (36 participants) who met each other for the first time in an immersive social VR platform. One of the general of findings of the current study is that the social VR environment could create a positive experience for older adults, supporting the limited number of previous studies that have evaluated older adult's responses to social VR environments [2, 3, 58, 63]. In addition to high ratings for self-report Engagement and for Willingness/Likelihood to Reconnect with a social partner encountered in the VR, majority of older participants in this study showed engagement in conversation and collaborative tasks, which in some cases ended up with exchanging contact information.

Four notable central findings from the current empirical study could inspire the future design of social VR for older adults. First, drawing on our observation and thematic analysis of the interaction of different pairs, we identified features of the VR content design that promoted and hindered social engagement among the older adult participants. We found that more relaxed and passive content, such as immersive video tours, were effective in promoting casual conversations. Often this led to discovery of shared personal interests and eventually long-term social connections. However, for users who are unenthusiastic about engaging with strangers in our case, the lack of interactivity could make this part of experience less engaging. In contrast, tasks in the interactive module (Module 3) "force" both users to be involved in order to reach the goals, guaranteeing conversations between them. While sufficient in quantity, these conversations were mostly about the tasks, and decisions to make, rather than more broadly social. While we found that most of users expected a high level of realism from the VR experience, we observed many scenarios showing a high degree of perceived presence among the participants, for example by expressing a fear of height in the VR, and an amused feeling of being "cold" in the water.

Second, our results suggest that the topics of the conversation were heavily influenced by the content of the VR. More "passive" contents such as the immersive videos tended to produce more far-reaching conversation, while more active and engaging content tended to encourage conversation focused on the immediate tasks. We noted that once participants had found a common interest or topic of discussion, this conversational thread tended to continue through the whole duration of the experiment. These findings are significant in that they indicate the diverse manner in which content may impact the direction of interactions—task-oriented content tended to produce more immediate engagement, but most participants seemed to prioritize the broader conversations that emerged around and beyond the tasks. This finding very much resembled Bartle's discussion of the high prevalence of "Socializers" in online games [5], those who treat the game as "merely a backdrop, a common ground." We also found task-oriented content, VR skills gaps among pairs, and familiarity with the VR content could lead to the dominancy of one of the pairs in social VR conversations. Pairing participants with the same level of familiarity with the technology and the VR content could help with a more balanced and equally engaging social VR experience for older adults.

Third, we identified characteristics among the older adults' users that contributed to social engagement in the VR environment, including proficiency with mobile devices and computers and shared interests between peers. Users with higher level of proficiency spent less time and effort on the interaction within the system, which allowed them to allocate more cognitive resources for social interaction, increasing their capabilities of social engagement. Shared interests between the two users also played an important role in terms of social engagement. As we described in the results, these interests themselves would encourage a great level of social engagement and endless conversation between our users, to a extend that nearly eliminated the need of an additional system to support it. We also found having both participants in a VR session with CI will results on a lot of conversation about the technical challenges and dependency on the moderator and less room for social interaction (P1). However, we did not observe any differences in social interaction behavior between pairs with one CI and one non-CI participant vs. pairs with both non-CI participants. Last and most fundamentally, not all the users were interested in making social connections with people they do not know. While it is nearly impossible for us to change that, pairing them with someone they already know, or giving some background information about their peer before the experiment may help with the situation. Some of our participants expressed a lack of interest in VR social interaction with strangers and mentioned that they would like to experience social VR with people they already know in real world. Additionally, while [4] suggested that social VR may "supported introverted users to express themselves", this was not found in our results where introvert users just chose not to talk to their peers.

Finally, we found many challenges caused by the design of the system, both in the hardware and software components. Similar to prior findings [16, 27, 30] we found relatively low levels of simulator sickness and task workload among older adults engaging with VR. However, the physical discomfort caused by the weight and tightness of the VR headset was mentioned as hardware challenges, which is congruent with other studies with older adults' participants [27, 30, 81]. In addition to more effective controls suitable for aging hands, a universal design approach to all aspects of VR equipment would be helpful to make the experience more accessible. It seemed clear in our findings that VR manufacturers have not given much attention to the ergonomic needs of older adults, even for very common issues such as eyeglass compatibility.

We observed a strong preference of the older adults for siting during the VR experience which cause a frequent mismatch between the physical and VR environment in height and a limited movement capability. As these technologies continue to grow and develop, we would encourage researchers to spend sufficient time on pre-experiment equipment training and to inform older participants about some of the common frustrations in this regard, so that they will not be unexpected. On the bright side, participants were often able to help each other to overcome the challenges, which in fact promoted the bonds between them. As on participant said, "We can get lost together."

## 5.1 Limitations and Future Work

One limitation of the study was lack of the simulation of the distance between users by decreasing the volume of their peers' speech [75], which was not included in our design due to concerns regarding hearing issues which are not uncommon among elders. In fact, we had to increase the volume, sometimes to levels that were uncomfortable loud for experimenters, at the beginning of the experiment to make sure that our participants could hear each other clearly. Applying room-scaled VR with locomotion control would also make the system more intuitive and immersive [65], which was not applied in this study.

The extent of moderation in the VR social activities is also an important variable. Our study used a relatively heavy moderation approach, particularly in the first two modules. Participants were assigned to a social partner based on matching schedules and did not have a selection choice in this matter, and the moderating researcher was heavily involved in initiating the introductory conversations. While a more open-ended social design might be preferable in some ways, it is important to think carefully about these parameters and to evaluate their impact, especially when working with potentially vulnerable populations such as individuals with CI. Obviously, there is a great deal of prospective research to carry out in this area to determine the optimal means of balancing user autonomy against potential harms (not to mention moderator fatigue) in the service of achieving the social program's stated goals.

While we placed the casual or "passive" video scenario first due to the nature of our tasks/goals, it would be interesting to explore the possibility of interlacing the two different types of scenarios, or at least reversing the order, allowing some initial bonds to be established in task-oriented interactions before letting the participants chat with each other without any concrete goal or topic. Furthermore, while we only included two scenarios in our study—one with high interactivity and clear goal, and another with low interactivity and no clear goal—it would be interesting to examine additional possibilities, such as scenarios with high environmental interactivity but no clearly specified goal (this is the case in many online games, such as *Minecraft*).

Another factor to consider when designing VR applications is the intended period of use. Our environment was designed, somewhat out of logistical necessity, as a very short-term experience, where users were able to meet new friends and make introductions. Such environments can potentially forge social connections that last after the experience, but duration is not built into the design. More persistent environments to which users can regularly return open up a much broader array of social design possibilities, such as long-term collaborative tasks and achievements or the creation of clubs/guilds. However, additional research is needed to evaluate how such long-term immersive worlds would affect older adults overall psychological and physical health.

**REFERENCES**

[1] Appel, L., Appel, E., Bogler, O., Wiseman, M., Cohen, L., Ein, N., Abrams, H.B. and Campos, J.L. 2020. Older adults with cognitive and/or physical impairments can benefit from immersive virtual reality experiences: a feasibility study. *Frontiers in medicine*. 6, (2020), 329.

[2] Baker, S., Kelly, R.M., Waycott, J., Carrasco, R., Bell, R., Joukhadar, Z., Hoang, T., Ozanne, E. and Vetere, F. 2021. School's Back: Scaffolding Reminiscence in Social Virtual Reality with Older Adults. *Proceedings of the ACM on Human-Computer Interaction*. 4, CSCW3 (2021), 1–25.

[3] Baker, S., Kelly, R.M., Waycott, J., Carrasco, R., Hoang, T., Batchelor, F., Ozanne, E., Dow, B., Warburton, J. and Vetere, F. 2019. Interrogating social virtual reality as a communication medium for older adults. *Proceedings of the ACM on Human-Computer Interaction*. 3, CSCW (2019), 1–24.

[4] Baker, S., Waycott, J. and Carrasco, R. 2021. Avatar-mediated communication in social vr: An in-depth exploration of older adult interaction in an emerging communication platform. *Conference on Human Factors in Computing Systems - Proceedings* (May 2021).

[5] Bartle, R. 1996. Hearts, clubs, diamonds, spades: Players who suit MUDs. *Journal of MUD research*. 1, 1 (1996), 19.

[6] Baumeister, R.F., Vohs, K.D. and Funder, D.C. 2007. Psychology as the science of self-reports and finger movements: Whatever happened to actual behavior? *Perspectives on psychological science*. 2, 4 (2007), 396–403.

[7] Benford, S., Greenhalgh, C., Rodden, T. and Pycock, J. 2001. Collaborative virtual environments. *Communications of the ACM*. 44, 7 (2001), 79–85.

[8]     Blackwell, L., Ellison, N., Elliott-Deflo, N. and Schwartz, R. 2019. Harassment in social virtual reality: Challenges for platform governance. *Proceedings of the ACM on Human-Computer Interaction*. 3, CSCW (2019), 1–25.

[9]     Boothby, E.J., Cooney, G., Sandstrom, G.M. and Clark, M.S. 2018. The liking gap in conversations: Do people like us more than we think? *Psychological science*. 29, 11 (2018), 1742–1756.

[10]    Bourdin, P., Sanahuja, J.M.T., Moya, C.C., Haggard, P. and Slater, M. 2013. Persuading people in a remote destination to sing by beaming there. *Proceedings of the 19th ACM Symposium on Virtual Reality Software and Technology* (2013), 123–132.

[11]    Braun, V. and Clarke, V. 2006. Using thematic analysis in psychology. *Qualitative research in psychology*. 3, 2 (2006), 77–101.

[12]    Brondi, R., Alem, L., Avveduto, G., Faita, C., Carrozzino, M., Tecchia, F. and Bergamasco, M. 2015. Evaluating the impact of highly immersive technologies and natural interaction on player engagement and flow experience in games. *International Conference on Entertainment Computing* (2015), 169–181.

[13]    Cotterell, N., Buffel, T. and Phillipson, C. 2018. Preventing social isolation in older people. *Maturitas*. 113, (2018), 80–84.

[14]    Coyle, H., Traynor, V. and Solowij, N. 2015. Computerized and virtual reality cognitive training for individuals at high risk of cognitive decline: systematic review of the literature. *The American Journal of Geriatric Psychiatry*. 23, 4 (2015), 335–359.

[15]    Darfler, M., Cruz-Garza, J.G. and Kalantari, S. 2022. An EEG-Based Investigation of the Effect of Perceived Observation on Visual Memory in Virtual Environments. *Brain Sciences*. 12, 2 (2022), 269.

[16]    Dermody, G., Whitehead, L., Wilson, G. and Glass, C. 2020. The role of virtual reality in improving health outcomes for community-dwelling older adults: systematic review. *Journal of medical internet research*. 22, 6 (2020), e17331.

[17]    Dickens, A.P., Richards, S.H., Greaves, C.J. and Campbell, J.L. 2011. Interventions targeting social isolation in older people: a systematic review. *BMC public health*. 11, 1 (2011), 1–22.

[18]    Dilanchian, A.T., Andringa, R. and Boot, W.R. 2021. A Pilot Study Exploring Age Differences in Presence, Workload, and Cybersickness in the Experience of Immersive Virtual Reality Environments. *Frontiers in Virtual Reality*. (2021), 129.

[19] Doucette, A., Gutwin, C., Mandryk, R.L., Nacenta, M. and Sharma, S. 2013. Sometimes when we touch: how arm embodiments change reaching and collaboration on digital tables. *Proceedings of the 2013 conference on Computer supported cooperative work* (2013), 193–202.

[20] Finstad, K. 2010. The usability metric for user experience. *Interacting with Computers*. 22, 5 (2010), 323–327.

[21] Freeman, G. and Maloney, D. 2021. Body, avatar, and me: The presentation and perception of self in social virtual reality. *Proceedings of the ACM on Human-Computer Interaction*. 4, CSCW3 (2021), 1–27.

[22] Ge, S., Zhu, Z., Wu, B. and McConnell, E.S. 2018. Technology-based cognitive training and rehabilitation interventions for individuals with mild cognitive impairment: a systematic review. *BMC geriatrics*. 18, 1 (2018), 1–19.

[23] Goll, J.C., Charlesworth, G., Scior, K. and Stott, J. 2015. Barriers to social participation among lonely older adults: The influence of social fears and identity. *PloS one*. 10, 2 (2015), e0116664.

[24] Hall, E.T. and Hall, E.T. 1966. *The hidden dimension*. Anchor.

[25] Hart, S.G. and Staveland, L.E. 1988. Development of NASA-TLX (Task Load Index): Results of Empirical and Theoretical Research. *Advances in Psychology*. Elsevier. 139–183.

[26] Herzog, A.R., Ofstedal, M.B. and Wheeler, L.M. 2002. Social engagement and its relationship to health. *Clinics in geriatric medicine*. 18, 3 (2002), 593–609.

[27] Huygelier, H., Schraepen, B., van Ee, R., vanden Abeele, V. and Gillebert, C.R. 2019. Acceptance of immersive head-mounted virtual reality in older adults. *Scientific Reports*. 9, 1 (2019), 4519. DOI:https://doi.org/10/ggwwtd.

[28] Jonas, M., Said, S., Yu, D., Aiello, C., Furlo, N. and Zytko, D. 2019. Towards a taxonomy of social vr application design. *Extended Abstracts of the Annual Symposium on Computer-Human Interaction in Play Companion Extended Abstracts* (2019), 437–444.

[29] Kalantari, S. 2019. A new method of human response testing to enhance the design process. *Proceedings of the Design Society: International Conference on Engineering Design* (2019), 1883–1892.

[30] Kalantari, S., Bill Xu, T., Mostafavi, A., Lee, A., Barankevich, R., Boot, W.R. and Czaja, S.J. 2022. Using a Nature-Based Virtual Reality Environment for Improving Mood States and

Cognitive Engagement in Older Adults: A Mixed-Method Feasibility Study. *Innovation in aging*. 6, 3 (2022), igac015.

[31] Kalantari, S. and Neo, J.R.J. 2020. Virtual environments for design research: lessons learned from use of fully immersive virtual reality in interior design research. *Journal of Interior Design*. 45, 3 (2020), 27–42.

[32] Kalantari, S., Rounds, J.D., Kan, J., Tripathi, V. and Cruz-Garza, J.G. 2021. Comparing physiological responses during cognitive tests in virtual environments vs. in identical real-world environments. *Scientific Reports*. 11, 1 (2021), 1–14.

[33] Kennedy, R.S., Lane, N.E., Berbaum, K.S. and Lilienthal, M.G. 1993. Simulator Sickness Questionnaire: An Enhanced Method for Quantifying Simulator Sickness. *The International Journal of Aviation Psychology*. 3, 3 (1993), 203–220. DOI:https://doi.org/10/bbh54v.

[34] Konijn, E.A., Utz, S., Tanis, M. and Barnes, S.B. 2008. *Mediated interpersonal communication*. Routledge New York, NY.

[35] Krueger, K.R., Wilson, R.S., Kamenetsky, J.M., Barnes, L.L., Bienias, J.L. and Bennett, D.A. 2009. Social engagement and cognitive function in old age. *Experimental Aging Research*. 35, 1 (Jan. 2009), 45–60. DOI:https://doi.org/10.1080/03610730802545028.

[36] Li, J., Kong, Y., Röggla, T., de Simone, F., Ananthanarayan, S., de Ridder, H., el Ali, A. and Cesar, P. 2019. Measuring and understanding photo sharing experiences in social virtual reality. *Proceedings of the 2019 CHI Conference on Human Factors in Computing Systems* (2019), 1–14.

[37] Li, J., Kong, Y., Röggla, T., de Simone, F., Ananthanarayan, S., de Ridder, H., el Ali, A. and Cesar, P. 2019. Measuring and understanding photo sharing experiences in social virtual reality. *Proceedings of the 2019 CHI Conference on Human Factors in Computing Systems* (2019), 1–14.

[38] Lundström, A. and Fernaeus, Y. 2019. The disappearing computer science in healthcare VR applications. *Proceedings of the Halfway to the Future Symposium 2019* (2019), 1–5.

[39] Malcolm, M., Frost, H. and Cowie, J. 2019. Loneliness and social isolation causal association with health-related lifestyle risk in older adults: a systematic review and meta-analysis protocol. *Systematic reviews*. 8, 1 (2019), 1–8.

[40] Maloney, D. and Freeman, G. 2020. Falling asleep together: What makes activities in social virtual reality meaningful to users. *Proceedings of the Annual Symposium on Computer-Human Interaction in Play* (2020), 510–521.

[41] Maloney, D., Freeman, G. and Robb, A. 2020. A Virtual Space for All: Exploring Children's Experience in Social Virtual Reality. *Proceedings of the Annual Symposium on Computer-Human Interaction in Play* (2020), 472–483.

[42] Maloney, D., Freeman, G. and Robb, A. 2020. It is complicated: Interacting with children in social virtual reality. *2020 IEEE Conference on Virtual Reality and 3D User Interfaces Abstracts and Workshops (VRW)* (2020), 343–347.

[43] Maloney, D., Freeman, G. and Wohn, D.Y. 2020. " Talking without a Voice" Understanding Non-verbal Communication in Social Virtual Reality. *Proceedings of the ACM on Human-Computer Interaction*. 4, CSCW2 (2020), 1–25.

[44] McGrath, J.E. 1984. *Groups: Interaction and performance*. Prentice-Hall Englewood Cliffs, NJ.

[45] McVeigh-Schultz, J., Kolesnichenko, A. and Isbister, K. 2019. Shaping pro-social interaction in VR: an emerging design framework. *Proceedings of the 2019 CHI Conference on Human Factors in Computing Systems* (2019), 1–12.

[46] McVeigh-Schultz, J., Márquez Segura, E., Merrill, N. and Isbister, K. 2018. What's It Mean to" Be Social" in VR? Mapping the Social VR Design Ecology. *Proceedings of the 2018 ACM Conference Companion Publication on Designing Interactive Systems* (2018), 289–294.

[47] Moustafa, F. and Steed, A. 2018. A longitudinal study of small group interaction in social virtual reality. *Proceedings of the 24th ACM Symposium on Virtual Reality Software and Technology* (2018), 1–10.

[48] Moustafa, F. and Steed, A. 2018. A longitudinal study of small group interaction in social virtual reality. *Proceedings of the 24th ACM Symposium on Virtual Reality Software and Technology* (2018), 1–10.

[49] Nasreddine, Z.S., Phillips, N.A., Bédirian, V., Charbonneau, S., Whitehead, V., Collin, I., Cummings, J.L. and Chertkow, H. 2005. The Montreal Cognitive Assessment, MoCA: A Brief Screening Tool For Mild Cognitive Impairment. *Journal of the American Geriatrics Society*. 53, 4 (2005), 695–699. DOI:https://doi.org/10/dmt678.

[50] Nicholson, N.R. 2012. A review of social isolation: an important but underassessed condition in older adults. *The journal of primary prevention*. 33, 2 (2012), 137–152.

[51] Nowak, K. and McGloin, R. 2014. The Influence of Peer Reviews on Source Credibility and Purchase Intention. *Societies*. 4, 4 (Dec. 2014), 689–705. DOI:https://doi.org/10.3390/soc4040689.

[52] Oh, C.S., Bailenson, J.N. and Welch, G.F. 2018. A systematic review of social presence: Definition, antecedents, and implications. *Frontiers in Robotics and AI*. (2018), 114.

[53] Pan, X. and Hamilton, A.F. de C. 2018. Why and how to use virtual reality to study human social interaction: The challenges of exploring a new research landscape. *British Journal of Psychology*. 109, 3 (2018), 395–417.

[54] Park, N.S. 2009. The relationship of social engagement to psychological well-being of older adults in assisted living facilities. *Journal of Applied Gerontology*. 28, 4 (2009), 461–481.

[55] Parsons, T.D., Gaggioli, A. and Riva, G. 2017. Virtual reality for research in social neuroscience. *Brain sciences*. 7, 4 (2017), 42.

[56] Rafnsson, S.B., Shankar, A. and Steptoe, A. 2015. Longitudinal influences of social network characteristics on subjective well-being of older adults: findings from the ELSA study. *Journal of Aging and Health*. 27, 5 (2015), 919–934.

[57] Ratan, R. and Hasler, B.S. 2014. Playing well with virtual classmates: relating avatar design to group satisfaction. *Proceedings of the 17th ACM conference on Computer supported cooperative work & social computing* (2014), 564–573.

[58] Roberts, A.R., de Schutter, B., Franks, K. and Radina, M.E. 2019. Older adults' experiences with audiovisual virtual reality: perceived usefulness and other factors influencing technology acceptance. *Clinical gerontologist*. 42, 1 (2019), 27–33.

[59] Röhr, S., Löbner, M., Gühne, U., Heser, K., Kleineidam, L., Pentzek, M., Fuchs, A., Eisele, M., Kaduszkiewicz, H. and König, H.-H. 2020. Changes in social network size are associated with cognitive changes in the oldest-old. *Frontiers in psychiatry*. 11, (2020), 330.

[60] Sanchez-Vives, M. v and Slater, M. 2005. From presence to consciousness through virtual reality. *Nature Reviews Neuroscience*. 6, 4 (2005), 332–339.

[61] Schuemie, M.J., van der Straaten, P., Krijn, M. and van der Mast, C.A.P.G. 2001. Research on presence in virtual reality: A survey. *CyberPsychology & Behavior*. 4, 2 (2001), 183–201.

[62] Siriaraya, P. and Ang, C.S. 2012. Characteristics and usage patterns of older people in a 3D online multi-user virtual environment. *Computers in Human Behavior*. 28, 5 (2012), 1873–1882.

[63] Siriaraya, P. and Ang, C.S. 2019. The Social Interaction Experiences of Older People in a 3D Virtual Environment. *Perspectives on Human-Computer Interaction Research with Older People*. Springer. 101–117.

[64]   Sra, M., Mottelson, A. and Maes, P. 2018. Your place and mine: Designing a shared VR experience for remotely located users. *Proceedings of the 2018 Designing Interactive Systems Conference* (2018), 85–97.

[65]   Sra, M., Mottelson, A. and Maes, P. 2018. Your place and mine: Designing a shared VR experience for remotely located users. *DIS 2018 - Proceedings of the 2018 Designing Interactive Systems Conference* (Jun. 2018), 85–98.

[66]   Tabbaa, L., Ang, C.S., Rose, V., Siriaraya, P., Stewart, I., Jenkins, K.G. and Matsangidou, M. 2019. Bring the outside in: providing accessible experiences through VR for people with dementia in locked psychiatric hospitals. *Proceedings of the 2019 CHI conference on human factors in computing systems* (2019), 1–15.

[67]   Tang, F., Chi, I. and Dong, X. 2017. The relationship of social engagement and social support with sense of community. *Journals of Gerontology Series A: Biomedical Sciences and Medical Sciences*. 72, suppl_1 (2017), S102–S107.

[68]   Thach, K.S., Lederman, R. and Waycott, J. 2020. How older adults respond to the use of virtual reality for enrichment: a systematic review. *32nd Australian Conference on Human-Computer Interaction* (2020), 303–313.

[69]   Välimäki, M., Hätönen, H.M., Lahti, M.E., Kurki, M., Hottinen, A., Metsäranta, K., Riihimäki, T. and Adams, C.E. 2014. Virtual reality for treatment compliance for people with serious mental illness. *Cochrane Database of Systematic Reviews*. 10 (2014).

[70]   Vorderer, P., Wirth, W., Gouveia, F.R., Biocca, F., Saari, T., Jäncke, L., Böcking, S., Schramm, H., Gysbers, A. and Hartmann, T. 2004. Mec spatial presence questionnaire. *Retrieved Sept*. 18, (2004), 2015.

[71]   Vozikaki, M., Linardakis, M., Micheli, K. and Philalithis, A. 2017. Activity participation and well-being among European adults aged 65 years and older. *Social Indicators Research*. 131, 2 (2017), 769–795.

[72]   Waycott, J., Kelly, R.M., Baker, S., Barbosa Neves, B., Thach, K.S. and Lederman, R. 2022. The Role of Staff in Facilitating Immersive Virtual Reality for Enrichment in Aged Care: An Ethic of Care Perspective. *Conference on Human Factors in Computing Systems - Proceedings* (Apr. 2022).

[73]   Wienrich, C., Schindler, K., Döllinger, N., Kock, S. and Traupe, O. 2018. Social presence and cooperation in large-scale multi-user virtual reality-the relevance of social interdependence for location-based environments. *2018 IEEE Conference on Virtual Reality and 3D User Interfaces (VR)* (2018), 207–214.

[74] Williamson, J., Li, J., Vinayagamoorthy, V., Shamma, D.A. and Cesar, P. 2021. Proxemics and social interactions in an instrumented virtual reality workshop. *Proceedings of the 2021 CHI Conference on Human Factors in Computing Systems* (2021), 1–13.

[75] Williamson, J.R., O'Hagan, J., Guerra-Gomez, J.A., Williamson, J.H., Cesar, P. and Shamma, D.A. 2022. Digital Proxemics: Designing Social and Collaborative Interaction in Virtual Environments. (Apr. 2022), 1–12.

[76] Winstead, V., Yost, E.A., Cotten, S.R., Berkowsky, R.W. and Anderson, W.A. 2014. The impact of activity interventions on the well-being of older adults in continuing care communities. *Journal of Applied Gerontology*. 33, 7 (2014), 888–911.

[77] Wrzus, C., Hänel, M., Wagner, J. and Neyer, F.J. 2013. Social network changes and life events across the life span: a meta-analysis. *Psychological bulletin*. 139, 1 (2013), 53.

[78] Wu, B. 2020. Social isolation and loneliness among older adults in the context of COVID-19: a global challenge. *Global health research and policy*. 5, 1 (2020), 1–3.

[79] Yassien, A., Elagroudy, P., Makled, E. and Abdennadher, S. 2020. A Design Space for Social Presence in VR. *ACM International Conference Proceeding Series* (Oct. 2020).

[80] Zamanifard, S. and Freeman, G. 2019. " The Togetherness that We Crave" Experiencing Social VR in Long Distance Relationships. *Conference Companion Publication of the 2019 on Computer Supported Cooperative Work and Social Computing* (2019), 438–442.

[81] Zhao, W., Baker, S. and Waycott, J. 2020. Challenges of Deploying VR in Aged Care: A Two-Phase Exploration Study. *ACM International Conference Proceeding Series* (Dec. 2020), 87–98.